\documentclass[runningheads]{svmult}

\usepackage{makeidx}   
\usepackage{graphicx}  
\usepackage{subeqnar}  
\usepackage{multicol}  
\usepackage{physprbb}  
\makeindex             

\def\arcmin{\ifmmode ^{\prime}\else$^{\prime}$\fi}
\def\arcsec{\ifmmode ^{\prime\prime}\else$^{\prime\prime}$\fi}
\def\approxlt{\mathrel{\hbox{\rlap{\lower.55ex \hbox {$\sim$}}
        \kern-.3em \raise.4ex \hbox{$<$}}}}
\def\approxgt{\mathrel{\hbox{\rlap{\lower.55ex \hbox {$\sim$}}
        \kern-.3em \raise.4ex \hbox{$>$}}}}

\begin{document}
\title*{X-ray Evidence for Supermassive Black Holes in Non-Active Galaxies: }
\subtitle{Detection of X-ray Flare Events, Interpreted 
           as Tidal Disruptions of Stars by SMBHs}
\toctitle{X-ray Evidence for SMBHs in Non-Active Galaxies
\protect\newline Detection of X-ray Flare Events, Interpreted 
 as Tidal Disruptions of Stars by SMBHs}
%
%
\titlerunning{X-ray Evidence for SMBHs in Non-Active Galaxies}
%
\author{Stefanie Komossa} 
\authorrunning{Stefanie Komossa}
%
%
\institute{Max-Planck-Institut f\"ur extraterrestrische Physik, Postfach 1312, 85741 Garching; skomossa@mpe.mpg.de}

\maketitle              

\vspace*{-7.8cm}
\begin{verbatim}
To appear in: LIGHTHOUSES OF THE UNIVERSE
              THE MOST LUMINOUS CELESTIAL OBJECTS AND THEIR USE FOR COSMOLOGY
              (held in Garching, August 2001)
              ESO Astrophysics Symposia, in press
\end{verbatim}
\vspace*{5.1cm}

\begin{abstract}
It has long been suggested that supermassive black holes in non-active galaxies
might be tracked down by occasional tidal disruptions of stars on nearly radial
orbits. A tidal disruption event would reveal itself by a luminous flare
of electromagnetic radiation.
Theorists argued that the convincing detection
of such a tidal disruption event would be the observation of an event
which fulfills the following three criteria: (1) the event should be of finite
duration (a `flare'), (2) it should be very luminous
(up to $L_{\rm max} \approx 10^{45}$ erg/s
in maximum), and (3) it should reside in a galaxy
that is otherwise perfectly {\em non}-active
(to be sure to exclude an upward fluctuation in
gaseous accretion rate of an {\em active} galaxy).
During the last few years, several X-ray flare events
were detected which match exactly
the above criteria. We therefore consider these events to be 
 excellent candidates for 
the occurrences of the theoretically predicted
tidal disruption flares.
In this contribution, we review the previous observations of
giant X-ray flares from normal galaxies, present new results
on these objects, critically discuss alternatives to the favored
outburst scenario, and report results
from our ongoing search for further tidal disruption
flares based on the {\sl ROSAT} all-sky survey database.

\end{abstract}

\section{Flares from tidally disrupted stars as probes for~the presence of SMBHs 
in {\itshape non-active} galaxies}
There is strong evidence for the presence of massive dark objects
at the centers of many galaxies. Does this hold for {\em all} galaxies ?
Questions of particular interest in the context of AGN evolution are:
what fraction of galaxies have passed through an active phase,
and how many now have non-accreting and hence unseen
supermassive black holes (SMBHs) at their centers
(e.g., Lynden-Bell 1969, Rees 1988)?
Several approaches were followed to study these questions.
A lot of effort has
concentrated on the determination of central object masses from studies of the {\sl dynamics of
stars and gas} in the nuclei of nearby galaxies.
Earlier (ground-based) evidence for central quiescent dark masses
in non-active galaxies
has been strengthened
by recent HST results
(see Kormendy \& Richstone 1995 for
a review).

Whereas the dynamics of stars and gas probe rather large
volumes, i.e., distances from the SMBH,
high-energy {\sl X-ray emission}
originates from the very vicinity of
the SMBH (see Komossa 2001 for a review).
In {\em active} galaxies, excellent evidence for the presence of SMBHs
is provided by the detection of luminous hard power-law like X-ray emission,
rapid variability, and the detection of relativistically 
broadened FeK$\alpha$ lines
(e.g., Tanaka et al. 1995).
How can we find {\em dormant} SMBHs in {\em non-active} galaxies ?
Lidskii \& Ozernoi (1979) and Rees (1988)
suggested to use the flare of electromagnetic radiation produced
when a star is tidally disrupted and accreted 
as a means to detect SMBHs in nearby non-active galaxies.

A star on a radial `loss-cone' orbit gets tidally disrupted after passing a
certain distance to the black hole (e.g., Hills 1975, Lidskii \& Ozernoi 1979,
Diener et al. 1997),
the tidal radius, given by
\begin{equation}
r_{\rm t} \simeq 7\,10^{12}\,({M_{\rm BH}\over {10^{6} M_\odot}})^{1 \over 3} ({M_{\rm *}\over M_\odot})^{-{1 \over 3}} {r_* \over r_\odot}~{\rm cm}\,. 
\end{equation}
The star is first heavily distorted, then disrupted.
About 50\%--90\% of the gaseous debris becomes unbound and is
lost from the system (e.g., Young et al. 1977, Ayal et al. 2000).
The rest will eventually be accreted by the black hole
(e.g., Cannizzo et al. 1990, Loeb \& Ulmer 1997).
The stellar material, first spread over a number of orbits,
quickly circularizes (e.g., Rees 1988, Cannizzo et al. 1990)
due to the action of strong
shocks when the most tightly bound matter interacts with
other parts of the stream (e.g., Kim et al. 1999).
Most orbital periods will then be within a few times
the period of the most tightly bound matter
(e.g., Evans \& Kochanek 1989).
A star will only be disrupted as long as its tidal radius
lies outside the Schwarzschild radius of the BH, else
it is swallowed as a whole (this happens for BH masses larger than
$\sim$10$^8$ M$_{\odot}$).{\footnote{Numerical simulations 
of the disruption process,
the stream-stream collision, the accretion phase,
and the depletion of loss-cone orbits and disruption rates have been studied in the
literature (e.g.,
Nolthenius \& Katz 1983,
Carter \& Luminet 1985, Evans \& Kochanek 1989,
Laguna et al. 1993, Diener et al. 1997, Ayal et al. 2000,
Kim et al. 1999,
Hills et al. 1975, Kato \& Hoshi 1978, Gurzadyan \& Ozernoi 1980, Cannizzo et al. 1990, Loeb \& Ulmer 1997,
DiStefano et al. 2001,
Frank \& Rees 1976, Magorrian \& Tremaine 1999; 
see Komossa \& Dahlem 2001 for more references, incl.  
a few observations of {\em active} galaxies that might be
related to tidal disruption events). Renzini et al. (1995; R95) reported
the detection of a UV flare from the (only mildly active) galaxy NGC\,4552.
The luminosity was several orders of magnitude weaker than what could have been expected
from a tidal disruption event. Tidal stripping of a star's atmosphere is
one possible explanation (R95). }}
More massive BHs may still disrupt or strip the atmospheres of giant stars.

\section{Tidal disruption flares from non-active galaxies: observational
evidence}

\subsection{Summary of X-ray and optical observations}

With the X-ray satellite {\sl ROSAT}, 
some rather unusual observations have been made in the last few
years: the detections of giant-amplitude, non-recurrent X-ray
outbursts from a handful of {\em optically non-active} galaxies,
starting with the case of NGC\,5905 (Bade et al. 1996, Komossa \& Bade 1999).
Based on the huge observed outburst luminosity,
the observations were interpreted in terms of tidal disruption events.
Below, we first give a brief review of the properties of all published
X-ray flaring non-active galaxies,
and then present results from a search for radio emission
from these galaxies.
{There are now four X-ray flaring `normal' galaxies}
{\mbox{(NGC\,5905, RXJ1242$-$1119, RXJ1624+7554, RXJ1420+}  5334{\footnote{The
X-ray position error circle of RXJ1420+53 contains a second galaxy for which a spectrum
is not yet available. Based on the galaxy's morphology, Greiner et al (2000) argue
that it is likely non-active}}), and a possible fifth
candidate (RXJ1331$-$3243), all of which show similar
properties (see Table 1 for a summary):
\begin{itemize}

\item
huge X-ray peak luminosity (up to $\sim 10^{44}$ erg/s),

\item
giant amplitude of variability (up to a factor $\sim$ 200),

\item
ultra-soft X-ray spectrum ($kT_{\rm bb} \simeq$ 0.04-0.1 keV), 

\item
absence of optical signs of Seyfert activity (the spectrum of NGC\,5905
is of HII-type; the other galaxies do not show any emission lines).

\end{itemize}

\begin{table*}[t]
\caption{Summary of the X-ray properties of the flaring non-active 
galaxies during outburst (NGC\,5905: Bade et al. 1996, Komossa \& Bade 1999,
RXJ1242$-$1119: Komossa \& Greiner 1999, RXJ1624+7554: Grupe et al. 1999,
RXJ1420+5334: Greiner et al. 2000; for first results on another
candidate see Reiprich \& Greiner 2001. Based on the position they report,
we refer to this source as RXJ1331$-$3243).
 $T_{\rm bb}$ denotes the black body
              temperature derived from a black body fit to the data
 (cold absorption was fixed to the Galactic value in the direction
 of the individual galaxies). $L_{\rm x,bb}$ gives the intrinsic luminosity in the
              (0.1--2.4) keV band, based on a black body fit.
 We note that this is a lower limit to the actual peak luminosity,
  since we most likely have not caught the sources exactly at maximum
light, since the spectrum may extend into the EUV, and since
we have conservatively assumed no X-ray absorption intrinsic to
  the galaxies. }
\begin{tabular}{cccc}
  \noalign{\smallskip}
  \hline
  \noalign{\smallskip}
galaxy name & ~~redshift $z$~~ & ~~$kT_{\rm bb}$ [keV] ~~& ~~$L_{\rm x,bb}$ [erg/s]~~  \\
  \noalign{\smallskip}
  \hline
  \hline
  \noalign{\smallskip}
  \noalign{\smallskip}
NGC\,5905 & 0.011 & 0.06 & 3 10$^{42}$$^*$  \\
  \noalign{\smallskip}
RXJ1242$-$1119 & 0.050 & 0.06 & 9 10$^{43}$~   \\
  \noalign{\smallskip}
RXJ1624+7554 & 0.064 & 0.097 & $\sim$ 10$^{44}$~  \\
  \noalign{\smallskip}
RXJ1420+5334 & 0.147 & 0.04 & 8 10$^{43}$~ \\
  \noalign{\smallskip}
RXJ1331$-$3243 & 0.051 &  &  \\
  \noalign{\smallskip}
\hline
\end{tabular}
\vskip0.15cm
  \noindent{\scriptsize $^{*}$Mean luminosity during the outburst; since the flux
 varied by a factor $\sim$3 during the observation, the peak luminosity is
higher. 
}
\end{table*}

\subsection{\bf Radio observations}

Radio observations are important for two reasons:
Firstly, they allow the search for a peculiar, optically hidden AGN at the center of
each flaring galaxy.
Secondly, radio emission could possibly be produced in relation to
the X-ray flare itself. 
We have performed a search for radio emission from the X-ray flaring
galaxies, based on the NRAO VLA Sky Survey (NVSS) catalogue (Condon et al. 1998)
at 1.4 GHz,
and the FIRST VLA sky survey at 1.5GHz (e.g., Becker et al. 1995).
With the exception of NGC\,5905, no flaring galaxy is radio-detected.
For NGC\,5905, several radio observations from the literature
are available, summarized in Fig. 1. The bulk of the radio emission
is extended, and NGC\,5905 does not show any peculiar radio
properties as compared to other similar spiral galaxies.  
Dedicated VLA radio observations of the nucleus of NGC\,5905, performed at a
frequency of 8.46 GHz several years after the X-ray outburst, 
provided an upper limit of 0.15 mJy for the presence of
a central point source (Komossa \& Dahlem 2001).

\begin{figure}[t]
\begin{center}
\includegraphics[width=.83\textwidth]{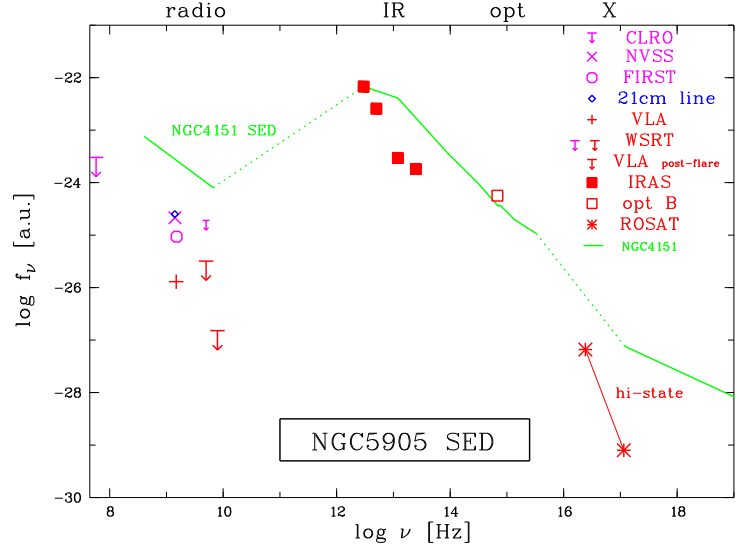}
\end{center}
\caption[]{Multiwavelength continuum spectrum of NGC\,5905 
(symbols represent data from Israel \& Mahoney 1990, van Moorsel 1982, Condon et al. 1998,
      Becker et al. 1995, Hummel et al. 1987, Brosch \& Krumm 1984, Komossa \& Dahlem 2001, NED,
      Bade et al. 1996, Komossa \& Bade 1999; the solid/dotted line
   corresponds to the continuum of the galaxy NGC\,4151, shown for comparison). 
   Note: data were taken with different
      aperture sizes and resolution, and at different times. }
\label{eps1}
\end{figure}

\section{\bf Outburst scenarios}

Most outburst scenarios do not survive
scrutiny (Komossa \& Bade 1999), because they
cannot account for the huge maximum luminosity (e.g.,
X-ray binaries within the galaxies, or a supernova in a dense environment),
are inconsistent
with the optical observations (gravitational lensing),
or predict
a different temporal behavior (X-ray afterglow of a Gamma-ray burst;
see, e.g., Fig. 2 of Bradt et al. 2001).
A critical discussion of AGN-related scenarios (presence of a dusty warm
absorber, or other absorption-related variablity), and   
why they are highly unlikely, is given by Komossa \& Dahlem (2001).

\subsection{\bf Tidal disruption model}

Except for some types of GRB-related emission mechanisms, the huge peak outburst
luminosity nearly inevitably calls for the presence of a SMBH.
This, in combination with the complete absence of any signs of AGN activity,
makes tidal disruption of a star by a SMBH
the most plausible outburst mechanism.

Intense electromagnetic radiation will be emitted
in three phases of the disruption and accretion process:
First, during the stream-stream collision when different parts
of the bound stellar debris interact with themselves (e.g., Rees 1988, Kim et al. 1999).
Secondly, radiation is emitted during the accretion of the stellar
material. Finally, the unbound stellar gas leaving the system
may shock the surrounding interstellar matter 
and cause intense emission, like in a supernova remnant (Khokhlov \& Melia 1996).

Although many details of the actual tidal disruption process are still unclear,
some basic predictions have been repeatedly made in the literature
how a tidal disruption event should reveal itself observationally:
(1) the event should be of finite
duration (a `flare'), (2) it should be very luminous
(up to $L_{\rm max} \approx 10^{45}$ erg/s
in maximum), and (3) it should reside in a galaxy
that is otherwise perfectly {\em non}-active
(to be sure to exclude an upward fluctuation in
gaseous accretion rate of an {\em active} galaxy).
All three predictions are fulfilled by the X-ray flaring galaxies;
particularly by NGC\,5905 and RXJ1242$-$1119, which are the two best-studied
cases so far.

In addition, we can do some further order of magnitude estimates and consistency checks.
The luminosity emitted if the black hole is accreting at its Eddington luminosity
can be estimated by 
\begin{equation}
L_{\rm edd} \simeq 1.3 \times 10^{38} M/M_{\odot} ~{\rm erg/s}\,.
\end{equation}
In case of NGC 5905, a BH mass of at least a few $\sim$$10^{4}$ M$_{\odot}$ would be
required to
produce the observed luminosity, and a higher mass if $L_{\rm x}$ was not observed
at its peak value.
For comparison, a BH mass of NGC\,5905 of $M_{\rm BH} \approx 10^{7} {\rm M}_\odot$
would be inferred, based on the correlation between bulge blue luminosity
and BH mass for spiral galaxies (Salucci et al. 2000),
or even up to a few 10$^{8} {\rm M}_\odot$ if we use 
the correlation reported mostly for ellipticals by Ferrarese \& Merritt (2001; 
their `sample A', their Fig.\,1).
For the other galaxies, using again $L_{\rm edd}$, we infer
BH masses reaching up to a few 10$^6$ M$_{\odot}$.
Alternative to a complete disruption event, the atmosphere of a giant star could have been
stripped. 
It is also interesting to note that NGC\,5905
possesses a complex bar structure (Friedli et al. 1996) which might
aid in the fueling process by disturbing the
stellar velocity fields.

In a simple black body approximation, the temperature
of the accretion disk scales with black hole mass 
as
\begin{equation}
T \simeq 8\,10^4 \,({M_{\rm BH}\over {M_\odot}})^{1 \over 12} ~{\rm K} ~~ ({\rm at}~ r_t), ~~~T \simeq 2\,10^7 \,({M_{\rm BH}\over {M_\odot}})^{-{1 \over 4}}~{\rm K} ~~ ({\rm at}~ 3\,r_S)\,.
\end{equation}
This gives $T_{\rm r_{tidal}} \simeq 3\,10^5$\,K, $T_{\rm 3r_S} \simeq 7\,10^5$\,K
for M=10$^6$\,M$_{\odot}$, where $r_{\rm S}$ is the  Schwarzschild radius.
Using black body fits of the X-ray flare spectra
we find 
temperatures in a similar range; $T_{\rm obs} \simeq$ (4-10)\,10$^5$ K. 
Like in AGN, X-ray powerlaw tails are possible. They might have
escaped detection during the observations, since weak,
or they may develop only after a certain time after the start
of the accretion phase.
We soon expect first results from a {\sl Chandra} and {\sl XMM} observation
of RXJ1242$-$1119, which will give valuable constraints on the post-flare evolution. 

The Eddington time scale for the accretion of the stellar material is given by 
\begin{equation}
t_{\rm edd} \simeq 4\,\eta_{0.1} (M_{\rm BH}/10^6M_{\odot}) (M_*/0.1M_{\odot}) ~{\rm yrs}\,.
\end{equation}
Uncertainties in estimating the total duration of the tidal disruption event
arise from questions like: how much material is actually accreted or expelled, does 
a strong wind develop, etc. The events are expected to last for months to years (e.g., Rees 1988). 
Observationally, the duration of the events was at least several days, followed by gaps in the observations. 
The source fluxes were then significantly down several years later (e.g., Fig.\,9 of Komossa \& Bade 1999). 

Finally, we note that the redshift distribution of the few sources observed
so far is consistent with the predicted tidal disruption rate,
in the sence that the events are sufficiently distant to define
a large volume of space, in which the detection of a few events would be expected.

\subsection{\bf Search for further X-ray flares}

We performed a search for further X-ray flaring activity 
using the sample of nearby galaxies of Ho et al. (1995) and
{\sl ROSAT} all-sky survey (Voges et al. 1999)
and archived pointed observations.
136 out of the 486 galaxies in the catalogue were observed at least twice with {\sl ROSAT}. 
We do not find another flaring normal galaxy in this sample, 
entirely consistent with the expected
tidal disruption rate of one event in at least $\sim$10$^4$ years per galaxy
(e.g., Magorrian \& Tremaine 1999).

\section{Outlook}
X-ray outbursts from non-active galaxies provide important information
on the presence of SMBHs in these galaxies,
and the link
between active and normal galaxies.
Future X-ray surveys,
like those planned with the {\sl LOBSTER} ISS X-ray all-sky monitor, {\sl ROSITA}
and {\sl MAXI}, 
will be valuable in finding more of these outstanding
sources.
Rapid follow-up observations at all wavelengths will then be important. 
In particular, X-ray observations with high spectral and temporal resolution might
open up a chance to probe the realm of strong gravity,
since the temporal evolution of the stellar debris will depend
on relativistic precession effects around the Kerr metric.

%


\begin{thebibliography}{8.}
\addcontentsline{toc}{section}{References}


\bibitem{}Ayal S., Livio M., Piran T., 2000, ApJ {\bf 545}, 772

\bibitem{}Bade N., Komossa S., Dahlem M., 1996, A\&A {\bf 309}, L35

\bibitem{}Becker R.H., White R.L., Helfand D.J., 1995, ApJ {\bf 450}, 559

\bibitem{}Bradt H., Levine A.M., Marshall F.E., et al. 2001, astro-ph/0108004

\bibitem{}Brosch N., Krumm N., 1984, A\&A {\bf 132}, 80

\bibitem{}Cannizzo J.K., Lee H.M., Goodman J., 1990, ApJ {\bf 351}, 38

\bibitem{}Carter B., Luminet J.P., 1985, MNRAS {\bf 212}, 23

\bibitem{}Condon J.J., Cotton W.D., Greissen E.W., et al., 1998, AJ {\bf 115}, 1693

\bibitem{}Diener P., Frolov V.P., Khokhlov A.M., et al.,  
             1997, ApJ {\bf 479}, 164

\bibitem{}Di\,Stefano R., Greiner R., Murray S., Garcia M., 2001, ApJL, in press

\bibitem{}Evans C.R., Kochanek C.S., 1989, ApJ {\bf 346}, L13

\bibitem{}Ferrarese L., Merrit D., 2001, ApJ, in press

\bibitem{}Frank J., Rees M.J., 1976, MNRAS {\bf 176}, 633

\bibitem{}Friedli D., Wozniak  H., Rieke M., Martinet L.,
            Bratschi P., 1996, A\&AS {\bf 118}, 461

\bibitem{}Greiner J., Schwarz R., Zharikov S., Orio M., 2000, A\&A {\bf 362}, L25

\bibitem{}Grupe D., Leighly K., Thomas H., 1999, A\&A {\bf 351}, L30

\bibitem{}Gurzadyan V.G., Ozernoi L.M., 1980, A\&A {\bf 86}, 315

\bibitem{}Hills J.G., 1975, Nature {\bf 254}, 295 

\bibitem{}Ho L.C., Filippenko A.V., Sargent W.L.W., 1995, ApJS {\bf 98}, 477

\bibitem{}Hummel E., van der Hulst J.M., Keel W.C., et al., 1987, A\&AS {\bf 70}, 517

\bibitem{}Israel F.P., Mahoney M.J., 1990, ApJ {\bf 352}, 30

\bibitem{}Kato M., Hoshi R., 1978, Prog. Theor. Phys. {\bf 60/6}, 1692

\bibitem{}Khokhlov A., Melia F., 1996, ApJ {\bf 457}, L61

\bibitem{}Kim S.S., Park M.-G., Lee H.M., 1999, ApJ {\bf 519}, 647

\bibitem{}Komossa S., 2001, in {\em IX. Marcel Grossmann Meeting
 on General Relativity, Gravitation and Relativistic Field Theories},
 V. Gurzadyan et al. (eds), World Scientific, in press [astro-ph/0101289]

\bibitem{}Komossa S., Bade N., 1999, A\&A {\bf 343}, 775

\bibitem{}Komossa S., Greiner J., 1999, A\&A {\bf 349}, L45

\bibitem{}Komossa S., Dahlem M., 2001, in {\em MAXI workshop on AGN variability}, in press
                [astro-ph/0106422] 

\bibitem{}Kormendy J., Richstone D.O., 1995, ARA\&A {\bf 33}, 581

\bibitem{}Laguna P., Miller W.A., Zurek W.H., Davies M.B., 1993, ApJ {\bf 410},
              L83

\bibitem{}Lidskii V.V., Ozernoi L.M., 1979, Sov. Astron. Lett. {\bf 5(1)}, 16

\bibitem{}Loeb A., Ulmer A., 1997, ApJ {\bf 489}, 573

\bibitem{}Lynden-Bell D., 1969, Nature {\bf 223}, 690

\bibitem{}Magorrian J., Tremaine S., 1999, MNRAS {\bf 309}, 447

\bibitem{}Nolthenius R.A., Katz J.I, 1983, ApJ {\bf 269}, 297

\bibitem{}Rees M.J., 1988, Nature {\bf 333}, 523

\bibitem{}Reiprich T., Greiner J., 2001, in 
          {\em ESO workshop on black holes}, 168

\bibitem{}  Renzini A.,
 Greggio L., Di Serego Alighieri S., et al.,
           1995, Nature {\bf 378}, 39

\bibitem{}Salucci P., Ratnam C., Monaco P., Danese L., 2000, MNRAS {\bf 317}, 488

\bibitem{}Tanaka  Y., Nandra K., Fabian A.C., et al., 1995, Nature {\bf 375}, 659

\bibitem{}van Moorsel G.A., 1982, A\&A {\bf 107}, 66

\bibitem{}Voges W., et al., 1999, A\&A {\bf 349}, 389

\bibitem{}Young P.,
                     Shields G.,
           Wheeler J.C., 1977, ApJ {\bf 212}, 367

\end{thebibliography}
\end{document}